\documentclass[12pt,a4paper]{article}
\usepackage{amsmath,amssymb}
\usepackage[dvips]{lscape,graphicx}
\usepackage{cite}
\usepackage{epsfig}
\begin{document}
\title{World - sheet Instantons in the Theory of the\\ QCD Strings}
\author{Georgy S. Iroshnikov\footnote{\large e-mail: irosh@orc.ru}}
\date{The Institute of Mathematical Sciences, Chennai 600 113, India;
\\ and \\ 
Moscow Institute of Physics and Technology, Institutskii proesd  9, 
Dolgoprudnyi, Moscow region, 141700, Russia.}
\maketitle

\begin{abstract}
The QCD string is manifested through the world-sheet instanton solutions 
which are responsible for confinement phenomenon and construction of 
$\theta$-vacua.
\end{abstract}
Keywords: QCD string, Instanton, Tension, $\beta$-function,
Semiclassics, Tachyon.

\section{Introduction }
This article is a continuation of my attempts to realize one possibility for 
extracting of the effective string dynamics from the strong-coupling regime 
of $SU(N)$ gauge theory at a large number N of colors. In past decade I had 
got the effective string picture with quantization of string action [1]. At 
this time it was looked as a some strange result. Now I understand that it 
is a manifestation of the world-sheet instantons which are responsible for 
confinement phenomenon and construction of $\theta$-vacuum.

Let us regard the nonperturbative calculation of hadron-field correlation 
functions
\begin{equation}
\label{eq1}
K(1,\dots,n)=\langle\Psi^+(x_1)\Psi(x_1)\cdots \Psi^+(x_n)\Psi(x_n)\rangle
\end{equation}
in the framework of the $SU(N)$ gauge field theory in the limiting case
\begin{equation}
\label{eq2}
N\gg Ne^2 \gg 1
\end{equation}
$$(e\; {\rm is\; the\; gauge\; coupling\; constant)}.$$ 
Here,
$$
\Psi^+(x)\Psi(x)\equiv\sum_{c=1}^N\Psi^+_c(x)\Psi_c(x)
$$
is a color-singlet operator of a composite point-like meson, and $\Psi_c(x)$ 
is a scalar (for the sake of simplicity) quark field, which is a spinor in 
the color space.

The hadron-field correlation function is
$$
K(1,\dots,n)= B^{-1}\int d\mu[A]D\Psi D\Psi^+e^{iS_{YM}
[A,\Psi,\Psi^+]}\times\{\Psi^+(x_1)\Psi(x_1)\cdots\Psi^+(x_n)\Psi(x_n)\},
$$
where $B$ is a normalization constant. Integrating with respect to $\Psi$ and 
discarding internal quark loops, which involve small factors of order $O(1/N)$,
we obtain the expression for the connected part of $K(1,\dots,n)$. It contains 
the Wilson loop and has the form
$$
K(1,\dots,n)\approx\int Dx_{\mu}(\gamma) D\lambda(\gamma)\langle 
C(\Gamma)\rangle_Ae^{i\oint \limits_{\Gamma}d\gamma\frac12(\dot 
x^2/\lambda+\lambda m_0^2)}.
$$
Here, $\gamma (0\le\gamma\le 1)$ is a variable specifying the parametrization 
of the contour $\Gamma$ that passes through the fixed points $x_1,\dots,x_n$ 
(see the figure);
$$
\dot x_{\mu}=\frac{dx_{\mu}}{d\gamma},\quad x_{\mu}(1)=x_{\mu}(0);
$$
$m_0$ is the bare quark mass; $\lambda(\gamma)$ is the one-dimensional metric 
on $\Gamma$; and 
\begin{gather}
\label{eq3}
\langle C(\Gamma)\rangle_A=B^{-1}\int d\mu[A]e^{iS_{YM}[A]}C(\Gamma),\\
\label{eq4}
C(\Gamma)=\text{\rm tr}\left[P\exp\left(ie\oint\limits_{\Gamma}
dx^{\mu}A_{\mu}\right)\right].
\end{gather}

We used the first quantization representation for propagator of scalar quark
\begin{align}
\Delta\left[x(\gamma_2);x(\gamma_1)\right]_{cc'}=&\int\limits_{x(\gamma_1)}
^{x(\gamma_2)}Dx_\mu(\gamma)\int D\lambda(\gamma)\notag\\
&\times e^{i\int\limits_{\gamma_1}^{\gamma_2}d\gamma\frac{1}{2}
(\dot x^2/\lambda+\lambda m_0^2)}\left\{P\exp\left[ie\int\limits_{\gamma_1}^
{\gamma_2}d\gamma\frac{dx_\mu}{d\gamma}A_\mu\right]\right\}_{cc'}\notag.
\end{align}

Let us introduce auxillary Grassman variables $\xi_c(\gamma)$ $(c=1,\dots,N)$ 
describing color spin of quarks $(\xi_c(\gamma)$ is a color spinor and a 
scalar with respect to the Lorentz group). Using these variables, we can 
recast $C(\Gamma)$ into the form of the path integral
\begin{equation}
\label{eq5}
C(\Gamma)=\int D\xi(\gamma)D\bar\xi(\gamma)\exp\{-iS[\xi,\bar\xi]-
\bar\xi_d(0)\xi_d(0)\}\xi_c(1)\bar\xi_c(0),
\end{equation}  
where
\begin{equation}
\label{eq6}
S[\xi,\bar\xi]=i\oint\limits_{\Gamma}d\gamma\bar\xi_c(\gamma)
\left[\frac{d}{d\gamma}-ieA_{\mu}\dot x^{\mu}\right]_{cd}\xi_d(\gamma).
\end{equation}
As a result of this transformation, the correlation function $K(1,\dots,n)$
assumes the form
\begin{align}
K(1,\dots,n)=&B^{-1}\int d\mu[A]Dx(\gamma)D\lambda(\gamma)D\xi(\gamma)
D\bar\xi(\gamma)\notag\\
\label{eq7}
&\times\xi_c(1)\bar\xi_c(0)\exp\{iS[A,\xi,\bar\xi,x]-\bar\xi_d(0)\xi_d(0)\},
\end{align}
where total action is
\begin{equation}
\label{eq8}
S[A,\xi,\bar\xi,x]=-\frac14\int d^4xG^a_{\mu\nu}G^{a,\mu,\nu} 
+\oint\limits_{\Gamma}d\gamma\left[\frac12(\dot x^2/\lambda
+\lambda m_0^2)+i\bar\xi D_{\gamma}\xi\right],
\end{equation}
\begin{equation}
\label{eq9}
G^a_{\mu\nu}=\partial_{\mu}A_{\nu}^a-\partial_{\nu}A_{\mu}^a
+ef^{abc}A_{\mu}^bA_{\nu}^c,
\end{equation}
and $D_{\gamma}$ is the covariant derivative with respect to the parameter 
${\gamma}$.

\section{The semi-classical $1/N$ approximation}
To calculate the correlation function (1), we will use the semi-classical 
$1/N$ expansion. It was shown in [2] that, in the limit of large $N$, the 
parameter $1/N$ plays the same role as Planck's constant $\hbar$ in the 
ordinary semi-classical approximation of quantum mechanics. In the leading 
order in $N$, we need take into account only  the contributions of planar 
gluon diagrams. This means that, in the path integral 
$\langle C(\Gamma)\rangle_A$ (3), it is necessary to carry out summation 
over the subset of gauge fields whose contributions are greatest in this 
approximation. We use topological arguments to single out this subset of 
fields. In order that the semi-classical $1/N$ approximation be valid, it 
is necessary to have a stable self-consistent extremal of the fields 
${\{\xi,A}\}$. Studies in the field of instanton and soliton physics showed 
that stable field configurations in nonlinear field theories correspond to 
topologically nontrivial solutions of classical equations of motion.

The variation of the action functional (8), that is, the condition 
${\delta S}/{\delta\bar\xi_c}=0$, gives the equation of motion
\begin{equation}
\label{10}
\frac{d\xi_c}{d\gamma}-ie(A_{\mu})_{cd}{\xi_d}\dot x^{\mu}=0.
\end{equation}
The formal solution of this equation is
\begin{equation}
\label{11}
\hat\xi_c(\gamma)=\left[P\exp\left(ie\int_0^{\gamma}d\gamma'\dot 
x^{\mu}A_{\mu}\right)\right]_{cd}{\xi_d(0)},
\end{equation}
where ${\xi_d(0)}$ is arbitrary. The ordered exponential function in (11) 
is defined on the contour ${\Gamma}$ and maps it into the $SU(N)$ group. 
This mapping is trivial for $N\ge 2$ because the homotopic group 
$\pi_1[SU(N)]=0$. The only exception arises when a quasi-Abelian field appears 
in expression (11); in this case, we are dealing with mapping of ${\Gamma}$ 
into  the subgroup ${U(1)}$ of the ${SU(N)}$ group. This mapping is nontrivial 
because we have $\pi_1[U(1)]= Z$, where $Z$ is the group of integers. Further, 
the field (11) will be a true classical solution of equation (10) if it is 
single-valued on the closed contour ${\Gamma}$. This leads to the quantization 
of the chromoelectric flux of the quasi-Abelian field through an arbitrary 
surface ${\Sigma}$ with the boundary ${\partial\Sigma}={\Gamma}$ [3]. This 
flux quantization can stabilize the field configuration 
${\{\hat\xi,\hat A}\}$ with respect to fluctuations if we are dealing with 
the full flux , and not with some part of the taken at random. Because the 
contour ${\Gamma}$ forms a one-dimensional boundary, this condition can be 
satisfied for a quasi-two-dimensional field of the form
\begin{equation}
\label{12}
A^a_{\mu}(x(z))\left.\vphantom{\frac{\partial x_{\mu}}{\partial z^i}}\right|_
{\Sigma}\equiv\left.\frac{\partial x_{\mu}}{\partial z^i}A^{a,i}(z)\right|_
{\Sigma},
\end{equation}
which is defined on the surface ${\Sigma}$ with the boundary 
${\partial\Sigma}={\Gamma}$. Here,
\begin{equation}
\label{13}
i=1,2;\qquad {\mu}=1,2,3,4;\qquad a=1,\dots,{N^2}-1;
\end{equation}
and the equation $x_{\mu}=x_{\mu}(z)$ determines
 the embedding of the surface ${\Sigma}$ into the flat space $R^{1,3}$ 
(or $R^4$ in the Euclidean formulation). Thus, the topologically nontrivial 
configuration $\{\xi[A],A\}$ is realized only in the subset of 
quasi-two-dimensional fields $A$ and the semi-classical $ 1/N $ calculations 
should be restricted to the evaluation of contributions from this subset of 
fields (12). In principle, there are contributions from other 
(four-dimensional) fields, but they do not give stable extremals in our 
problem. Therefore, these contributions cannot be calculated with the aid 
of the semi-classical expansion, and some other methods must be designed to 
accomplish these ends. However, there is independent evidence that, in the 
leading order of the strong - coupling approximation, the string tension is 
dominated by contributions from two-dimensional fields.This statement follows 
from a comparison of the string tension that is calculated in the Hamiltonian 
formulation of the lattice gauge theory [4] with the result obtained by 
computing Wilson loop in the Yang-Mills theory on arbitrary (two- dimensional) 
surfaces [5]. These results also coincide with the string tension (26) 
obtained 
in this study. The coincidence of these results shows that the semi-classical 
$1/N $ approximation includes the leading contributions to hadron correlators 
in the strong - coupling regime. 
So, if we take into account only contributions of fields type (12), the 
calculation of the expectation value in (7) is reduced to integration with 
respect to the two-dimensional field $ [A^a_i(z)]_{\Sigma}$ for a fixed 
surface ${\Sigma}$ and subsequent summation over surfaces ${\Sigma}$. 

Going over to the new variables, we obtain the effective two-dimensional 
action for each surface ${\Sigma}$ [6].
\begin{align}
\label{14}
S^{eff}_{\Sigma}[A,{\xi},x,{\lambda}]=& - \frac14\int\limits_{\Sigma}d^2z(-h)^
{\frac12}h^{il}h^{kn}G^a_{ik}G^a_{ln}\notag\\ 
&+\oint\limits_{\partial\Sigma}d{\gamma}\frac12( \dot x^2/\lambda+\lambda 
m_0^2)+i\oint\limits_{\partial\Sigma}d{\gamma}
\bar\xi_c\left(\frac{d}{d\gamma}- i{\varepsilon}A_i {\dot z}^i\right)_{cd}
{\xi_d},
\end{align}
\begin{equation}
\label{15}
G^a_{ik}=\partial_iA^a_k-\partial_kA^a_i+{\varepsilon}f^{abc}A^b_iA^c_k.
\end{equation}
Here, ${\varepsilon}=e/{\delta}$ is the two-dimensional charge, ${\delta}$ is 
the UV regulator with the dimension of length, $h=det(h_{ik})$, and
$$
h_{ik}=\frac{\partial x_{\mu}}{\partial z^i}\frac{\partial x^{\mu}}{\partial 
z^k}
$$
is the metric induced on the surface $\Sigma$ by embedding 
$x_{\mu}=x_{\mu}(z)$.

\section{World-sheet instantons as a solutions of the\\ classical equations}

The condition $\delta S^{eff}/\delta A^a_i=0 $ gives the equation of motion
\begin{equation}
\label{16}
\partial_i(\sqrt{-h}G^{a,ik})-{\varepsilon}f^{abc}\sqrt{-h}G^{b,ik}A^c_i=0
\end{equation}
for the field $A^{cl}(z)\equiv\hat A(z)\ $ on the surface $\Sigma$ and the 
boundary condition
\begin{equation}
\label{17}
\sqrt{-h}G^{a,ik}(z({\gamma}))e_{si}{\dot z}^s = {\varepsilon}T^a({\gamma}) 
{\dot z}^k
\end{equation}
on ${\delta}{\Sigma}={\Gamma}$.
The boundary condition takes into account the presence of quarks at the 
boundary of the surface. Here, $ e_{si}$ is the antisymmetric unit tensor, and
\begin{equation}
T^a(\gamma)=\bar\xi_c(\gamma)\left(\frac{\lambda^a}{2}\right)_{cd}\xi_d(\gamma)
\label{18}
\end{equation}
 is the operator of the quark color spin. By virtue equation (10), the 
operator $ T^a(\gamma) $ is a covariantly constant quantity.
Equation (16), together with the boundary condition (17), have a solution 
of the form
\begin{equation}
G^{a, ik}(z)=\varepsilon\left(e^{ik}/\sqrt{-h(z)}\right)I^a(z),
\label{19}\end{equation}
provided that $I^a(z)$ satisfies the equation
\begin{equation}
D_i^{ab}I^b(z)=0.
\label{20}\end{equation}
Tensor (19) also obeys the two-dimensional Bianchi identity. In view of 
boundary condition (17), the square of the vector $ I^a $ equals the square of 
the quark color spin; that is 
\begin{equation}
I^2=I^aI^a=(N^2-1)/2N.
\label{21}\end{equation}
In a special gauge where 
\begin{equation}
I^a={\rm const},
\label{22}
\end{equation}
the potential $\hat A$ corresponding to (19) has the form
\begin{align}
\hat A_i^a(z)=&I^aa_i(z)/\varepsilon,\notag\\
a_i(z)=&\frac{e_{ik}}{2}\sqrt{-\hat h}\hat h^{kl}\partial_l\ln\sqrt{-\hat h}
\label{23}
\end{align}
where $\hat h $ is the metric of the constant curvature
\begin{equation}
R=-2\varepsilon^2
\label{24}
\end{equation}
on $\Sigma$.
Substituting the self-consistent configuration $\{\hat A, \hat\xi\}$ into 
the action (14) and going over to the Euclidean space, we reduce $ S^{eff}$ 
to the following form of the string action with constraints (only in 
Euclidean space above action has a right sign):
\begin{align}
S_\Sigma^{eff}[x,\lambda]=&k_0\int\limits_{\Sigma}d^2z\sqrt{h(x(z))}\notag\\
&+\frac{1}{2}\oint\limits_{\Gamma=\partial\Sigma}d\gamma\left[
\dot x^2(z(\gamma))/\lambda(\gamma)+\lambda(\gamma)m_0^2\right]+
{\rm constraints},
\label{25}
\end{align}
where  
$$
\dot x_\mu=\frac{dx_\mu}{d\gamma}=\frac{dx_\mu}{dz^i}\frac{dz^i}{d\gamma}.
$$
The bare string tension is given by
\begin{equation}
k_0=\frac{e^2}{2\delta^2}\left(\frac{N^2-1}{2N}\right),\quad e^2=e_0^2/N,\quad 
e/\delta\equiv\varepsilon.
\label{26}
\end{equation}
 (The string tension $k$ is renormalized as the result of subsequent summation 
over surfaces. In this procedure, the parameter $\delta$ is made to tend to 
zero; this is equivalent to the removal of the cut-off and the introduction 
of the normalization point.) In the limit $N \to \infty$, $k_0$ is independent 
of $N$; that is, we have $S^{eff}[\hat A,\hat\xi]\sim O(1)$.
The constraints that enter in formula (25) ensure that our solutions will be 
the world-sheet instantons: a topologically nontrivial embedding of the world 
sheet in target space.
\section{Constraints}
The first constraint imposes condition of quantization on the string action; 
that is,
\begin{equation}
k_0\int\limits_{\Sigma}d^2z\sqrt{h}=\pi|Q|,\quad  Q=0,\pm 1, \pm 2, \dots
\label{27}
\end{equation}
This condition is a corollary of the quantization of the chromoelectric flux 
on the world sheet $\Sigma$. Flux quantization follows from the requirement 
that the classical solution $\hat\xi_c(\gamma)$ (11) defined on the closed 
contour ${\Gamma} = \partial\Sigma$ be single-valued. The index $Q$ in (27) 
determines the total increment of the phase of $\hat\xi(\gamma)$ upon the 
circumvention along the contour $\Gamma$. This index has the gauge invariant 
representation
\begin{equation}
Q=\frac{\varepsilon}{4\pi}\int\limits_{\Sigma}d^2ze^{ik}I^a(z)G^a_{ik}(z)
\label{28}
\end{equation}
and characterizes various topological sectors. In the special gauge (22), 
(23), it assumes the simple form
\begin{equation}
Q=\frac{I^2}{2\pi}\oint\limits_{\Gamma}dz^ia_i.
\label{29}
\end{equation}

Furthermore, it can be shown [3] that the gauge component of the action(14) 
is bounded from below, namely,
\begin{equation}
S_{YM}[A]\ge \pi|Q|.
\label{30}
\end{equation}
As a result, the string field configuration $ \{\hat A,\hat\xi\} $ turns out 
to be stable to small fluctuations of the gauge field that refer to a definite 
sector specified by $Q$.

The second constraint is associated with condition (24), which requires that 
the scalar curvature be constant. If condition (24) were not imposed, it 
would be possible to contract the boundary $\partial\Sigma$ of the surface 
$\Sigma$ to zero by deforming the surface without changing the surface area 
$A(\Sigma)$. This would destroy the topological classification (27) with 
$Q \neq  0$. Moreover, condition (24) ensures the $1/N$ - suppression of 
Gaussian fluctuations $\delta A$ [7]. The solution of (23) expressed in terms 
of metric $\hat h$ actually associates the spin connection on the surface 
$\Sigma$ with the vector potential $a_i(z)$ or, which is equivalent, 
associates the curvature tensor with the strength tensor
$F_{ik}=\partial_{[i} a_{k]}$ on $\Sigma$. Thus, in the approach developed 
here, the geometry of the string world sheet is determined by the dynamics of 
the gauge field on the same sheet. So, we have the manifestation of 
world-sheet instantons - topologically nontrivial embedding of the string 
sheet in outer space.

\section{The partition function and correlation functions}

In view of the quantization condition (27), the partition function
\begin{equation}
Z=\int Dx_\mu(z)D\lambda(\gamma)\exp\{-S[x(z),\lambda(\gamma)\}
\label{31}
\end{equation}
is decomposed into the sum of contributions from all topological sectors; 
that is, we have
\begin{align}
Z=&\sum\limits_{|Q|=0}^{\infty}Z_{|Q|}=\sum\limits_{|Q|}
\left(Z_{Q^+}+Z_{Q^-}\right),\notag\\
Q^{\pm}=&\pm|Q|.
\label{32}
\end{align}

The calculation of $Z_Q $ is carried out with the aid of the relation
\begin{align}
\int Dx_\mu\exp\left[-k_0\int\limits_{\Sigma}d^2z\sqrt{h}\right]\circeq&
\int Dx_\mu Dg_{ab}\notag\\
&\times \exp\left(-\frac{k_0}{2}\int\limits_{\Sigma}d^2z\sqrt{g}g^{ab}
\partial_ax_\mu\partial_bx_\mu\right),
\label{33}
\end{align}
which holds only in the leading approximation of the of steepest descent.
(The symbol $\circeq$ means that the integral with respect to $g_{ab}$ 
equals the value of the integrand taken at the saddle point.)

The calculation of the partition function $Z$ by means of integration the 
conformal anomaly [8] with allowance for the constraints showed [9] that 
the partition function $Z_Q$ is expressed only in terms of the Euler 
characteristic $\chi$ of the world sheet, that is,
\begin{equation}
Z_{|Q|}\sim\exp\left[-D(\chi/6+1/3)|Q|\right].
\label{34}
\end{equation}
This means that, in the approximation used here, the $SU(N)$ gauge theory is 
reduced to a topological theory. The conformal anomaly  proportional to the 
Liouville action functional is neutralized by the condition 
$R = {\rm constant}$: that is, there is no critical dimension $D=26$.

In the calculation of $Z$ (31), there arise divergences proportional to the 
surface area and perimeter of the world sheet. This enable us to renormalize 
the string tension $k$ and the quark mass $m_q$ [9].
The renormalized mass of a quark determines the geodesic curvature of the 
boundary $\partial\Sigma$; that is, $\kappa_{\hat g}=m/4$.
In particular there is relation
\begin{equation}
\kappa^2_{\hat g}=-R/2
\label{35}
\end{equation}
which is a characteristic of the Beltrami pseudo-sphere, that is, our 
instanton wraps around noncontractible surface in target space that is 
pseudo-sphere. We also have the relation $k\sim m^2\sim R$ .

Thus, there is only one independent dimensional parameter, namely,
the renormalized string tension
\begin{equation}
k\equiv\frac{1}{2\pi\alpha'}=\frac{e^2(a)}{2a^2}\left(\frac{N^2-1}{2N}\right).
\label{36}
\end{equation}
Its value can be determined by analyzing experimental data that concern 
distances $a$ on the order of the confinement radius ($a\sim r_{conf}$). Here, 
$e(a)$ is the running coupling constant, and $a$ is the normalization point. 
In the nonperturbative analysis of a field theory that admits quark 
confinement, it is natural to use the renormalization scheme in which the 
string tension $k$ is fixed:
\begin{equation}
\frac{dk}{da}=0.
\label{37}
\end{equation}
(The quantity $\sqrt k$ plays a role here similar to that of the dimensional 
perturbative parameter $\Lambda_{QCD}$; these two quantities are related by a 
linear equation.) From condition (37) and from equation (36), it follows that
 the Gell-Mann-Low function has the form 
\begin{equation}
\beta(e)\equiv -a\frac{de}{da}=-e(a).
\label{38}
\end{equation}
This expression coincides with the first term of the expansion of the 
$\beta$-function in inverse powers of the charge $e$ in the strong-coupling 
approximation of the Hamiltonian formulation of the lattice gauge theory [4]. 
This formulation is convenient because it permits the separation of 
contributions from electric and magnetic color fields. In the strong-coupling 
approximation, the contribution of the electric field is dominant. At 
distances $a\ll\ r_{conf}$, where the charge is small, it is necessary to 
take into account the contribution of the magnetic field as well. Numerical 
calculations showed [4] that, in this case, expression (38) goes over 
smoothly to the standard formulas
\begin{align}
\beta(e)&=-b_0e^3-b_1e^5\dots,\notag\\
b_0&=\frac{11}{48\pi^2}N,\quad  
b_1=\frac{34}{3}\left(\frac{N}{16\pi^2}\right)^2
\label{38}
\end{align}
\label{39}
for $\beta (e)$ in the weak-coupling approximation.

The final stage of the calculation of hadron-field correlation functions is 
as follows [10]. In the approximation of scalar quarks, the above functions 
are reduced to Koba-Nielsen dual resonance amplitudes in each topological 
sector. It turned out that the above-listed distinctions of the chromoelectric 
string from the standard model manifest themselves mainly in the calculation 
of the partition function and drop out,to a considerable extent, of final 
expressions for correlation functions. However, there is still a tachyon in 
the ground state in each topological $Q$-sector. An infinite number of 
degenerate $Q$-sector enables us to consider a new possibility for solving 
the long-standing tachyon problem in dual resonance models. Namely, the 
introduction of the $\theta$-vacuum,which is a superposition of these sectors,
is expected to result in a shift in the mass spectrum and lead to the 
elimination of the tachyon state from the theory.

\section{Homotopical classification of backgrounds,\\ $\theta$-vacuum and 
tachyon mass problem}

The solution of the equation of motion (10) according to (11), (23) and [6]
has a form
\begin{equation}
\xi_c(\gamma)=e^{-i\phi(\gamma)}\xi_c(0)
\label{40}
\end{equation}
where the phase
\begin{equation}
\phi(\gamma)=I^2\int\limits_0^\gamma d\gamma'\frac{dz^i}{d\gamma'}a_i(z)=
I^2\int\limits_0^\gamma d\gamma' a_{\gamma'}
\label{41}
\end{equation}
does not depend on the index $c= 1,\dots ,N$ this being a consequence of the 
spontaneous breaking of the $SU(N)$ symmetry to $U(1)$. The phase 
$\phi(\gamma)$ forms mappings $S_1 \to  S_1$, characterized by a winding 
number $Q$ (28), (29). Then we can construct the backgrounds (pure gauge) at 
the boundary $\Gamma$ for each sectors $Q$
\begin{equation}
a_\gamma=i\Lambda_Q^{-1}\frac{\partial}{\partial\gamma}\Lambda_Q=
\frac{2\pi Q}{I^2},
\label{42}
\end{equation}
where
\begin{equation}
\Lambda_Q(\gamma)=e^{-i2\pi Q\gamma/I^2},
\label{43}
\end{equation}
and
\begin{equation}
\phi(\gamma=1)\equiv I^2\oint\limits_\Gamma d\gamma a_\gamma= 2\pi Q.
\label{44}
\end{equation}
Because the our world-sheet instanton has topological index $Q$=1, it can 
makes the communication between next backgrounds. Anticipating that some 
tunnelling does take place, the correct vacua will be linear combinations of 
the topological vacua and are given by
\begin{equation}
|\theta\rangle=\sum\limits_{Q=-\infty}^\infty e^{i\theta Q}|Q\rangle.
\label{45}
\end{equation}
 It is equivalent to add the $\theta$ -term to the action [11]
\begin{equation}
S_\theta=S_0+i\theta Q.
\label{46}
\end{equation}
Each tachyon for sector $Q$ has the statistical weight is equal to 
(for $D=4$) [9]
\begin{equation}
(\beta A_{eff})^{|Q|}\frac{e^{-\frac{4}{3}|Q|}.e^{i\theta Q}}{|Q|!},\quad 
S_0=\frac{4}{3}|Q|
\label{47}
\end{equation}
$$
{\rm (The\; Euler\; characteristic}\;\; \chi=0\;\; {\rm for\; pseudo-sphere).} 
$$

The pre-exponential factor taking into account the contribution from Gaussian 
fluctuations of gauge field around the instanton solution. According to 
[7] this factor is proportional to effective instanton area 
$A_{eff}=2{\pi} a^2$, where $a$-radius Beltrami pseudo-sphere. 
The $   |Q|!   $ arise because of indistinguishability of instantons.

Let us regard the contribution of one tachyon pole in the dual resonance 
amplitude $A(s,t)$. It is equal to
\begin{equation}
\frac{1/\alpha'}{m_0^2-s},\quad  m_0^2< 0.
\label{48}
\end{equation}
Then one makes the transformation
\begin{equation}
\int\limits_{-\infty}^{\infty}\frac{e^{-isA_M}}{m_0^2-s-i\varepsilon}
=e^{-im_0^2A_M}
\to e^{-m_0^2A_E}
\label{49}
\end{equation}
where we have came from Minkovski area space $A_M$ to Euclidean area 
$A_E=A_{eff}$. If we multiply this contribution by factor (47) and make the 
sum over all topological sectors with $n$ instantons and $\bar n$ 
anti-instantons, we will get to result
\begin{align}
&\frac{1}{\alpha'}e^{-m_0^2A_{eff}}\sum\limits_{n,\bar n=0}^{\infty}
\frac{\left(\beta A_{eff}e^{-S_0+i\theta}\right)^n}{n!}
\frac{\left(\beta A_{eff}e^{-S_0-i\theta}\right)^{\bar n}}{\bar n!}\notag\\
&=\frac{e^{-m_0^2A_{eff}}}{\alpha'}\exp\left\{
e^{-S_0}\beta A_{eff}2\cos\theta\right\}
=\frac{1}{\alpha'}e^{-m^2(\theta)A_{eff}},
\label{50}
\end{align} 
where
\begin{equation}
m^2(\theta)=m_0^2-2\beta e^{-S_0}\cos\theta.
\label{51}
\end{equation}
So, we have got the shift of the tachyon mass. 

Let us stress also that $\theta$-term in (46) does not violate the parity 
symmetry because contains only the internal two-dimensional potential. Also 
there is not any violation of the charge conjugation symmetry, because 
$\theta$-term contains charge squared according to (28), (19).

\section{Conclusion}
In the leading order of the semi-classical $1/N$ expansion, the partition 
function and correlation functions are apparently dominated by the 
world-sheet instantons solutions. This ensures the string picture with 
quark confinement and introducing of the $\theta$-vacua.

\section*{Acknowledgements}
I am grateful to the Institute of Mathematical Sciences (Chennai, India) 
for its hospitality, to Professor N.D. Hari Dass for useful discussions, to 
Dr. Chitta Ranjan Das and Dr. Ioulia Baoulina for help in the preparation of 
this manuscript.

\newpage
\begin{figure}
\begin{center}
{\includegraphics[height=180mm,keepaspectratio=true]{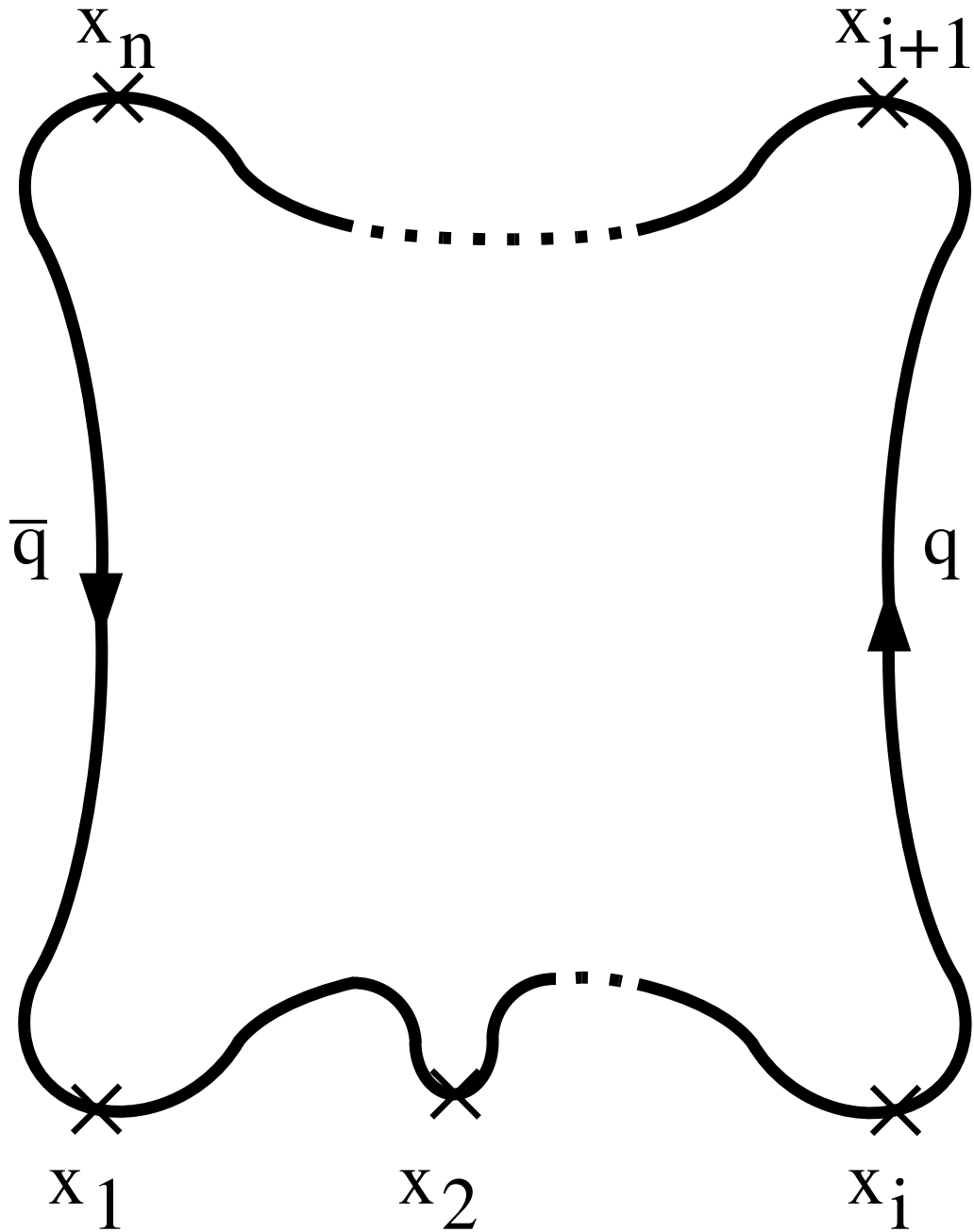}}
\caption{The contour $\Gamma$ in 4-dim, external space corresponding to the
connected part of the correlator $K(x_1,\dots,x_n)$.}
\end{center}
\label{fig1}
\end{figure}
\end{document}